\documentclass[pra,twocolumn,showpacs]{revtex4}
\usepackage{amssymb}

\begin{document}

\title{Necessary and Sufficient Conditions for the Trumping Relation}
\author{S. Turgut}
\email{sturgut@metu.edu.tr}
\affiliation{ Department of Physics,
Middle East Technical University, 06531, ANKARA, TURKEY}

%\date{\today}

\begin{abstract}
Entanglement catalysis allows one to convert certain entangled
states into others by the temporary involvement of another
entangled state (so-called catalyst), where after the conversion
the catalyst is returned to the same state. For bipartite pure
entangled states that can be transformed in this way with unit
probability, the respective Schmidt coefficients are said to
satisfy the trumping relation, a mathematical relation which is an
extension of the majorization relation. This article provides all
necessary and sufficient conditions for the trumping relation in
terms of the Schmidt coefficients. The coefficients should satisfy
strict inequalities for the entropy of entanglement and for power
means excluding the special power 1.
\end{abstract}



\maketitle

\section{Introduction}

An important problem in quantum information theory is to
understand the conditions for transforming a given entangled state
into another desired state by using only local quantum operations
assisted with classical communication (LOCC). Significant
development has been achieved for the case of pure bipartite
states. In the asymptotic limit, where an infinite number of
copies of entangled pairs are to be transformed into each other,
it is found that the conversion is possible with the probability
of success approaching unity as long as the entropy of
entanglement does not increase.\cite{Bennett1}

However, away from the asymptotic limit, where a single copy of a
given state is to be transformed into another given state, such a
simple conversion criterion cannot be found and investigations
have unearthed a deep connection of the problem to the
mathematical theory of majorization.\cite{NielsenMaj} For two
sequences of $n$ real numbers $x=(x_1,x_2,\ldots,x_n)$ and
$y=(y_1,y_2,\ldots,y_n)$, we say that $x$ is \emph{majorized} by
$y$ (written $x\prec y$) when
\begin{equation}
 x_1^\uparrow+x_2^\uparrow+\cdots+x_m^\uparrow \ge y_1^\uparrow+y_2^\uparrow+\cdots+y_m^\uparrow
\label{eq:Major}
\end{equation}
for $m=1,2,\ldots,n-1$ and the sequences have the same sum ($\sum
x_i=\sum y_i$). Here, $x^\uparrow$ represents the sequence $x$
when its elements are arranged in non-decreasing order
($x_1^\uparrow\le x_2^\uparrow\le\cdots\le x_n^\uparrow$) and
similarly for $y^\uparrow$. Nielsen has shown that for two given
states having the Schmidt forms
\begin{eqnarray}
\vert\psi\rangle &=& \sum_{i=1}^n \sqrt{x_i}\vert i_A\otimes i_B\rangle\quad,\\
\vert\phi\rangle &=& \sum_{i=1}^n \sqrt{y_i} \vert i_A^\prime \otimes i_B^\prime\rangle\quad,
\end{eqnarray}
where $x$ and $y$ are the respective Schmidt coefficients ($\sum
x_i=\sum y_i=1$), the state $\vert\psi\rangle$ can be converted
into $\vert\phi\rangle$ by LOCC with unit probability of success
if and only if $x\prec y$.\cite{Nielsen} Subsequently, Vidal
extended this result to probabilistic transformations where the
probability of success is related to the violation of the
majorization inequalities in (\ref{eq:Major}).\cite{Vidal}

Soon afterwards, Jonathan and Plenio have demonstrated an
interesting effect that is termed as catalysis or entanglement
assisted local transformation.\cite{JonathanPlenio} There are some
cases where $\vert\psi\rangle$ cannot be converted into
$\vert\phi\rangle$ with certainty (conversion is possible only
with a probability less than 1), but with the involvement of
another entangled pair (a catalyst), the conversion is made
possible. In other words, if $\vert\chi\rangle=\sum_{\ell=1}^N
\sqrt{c_\ell}\vert \ell_A\otimes\ell_B\rangle$ is the state of the
catalyst, then $\vert\psi\rangle\otimes\vert\chi\rangle$ can be
converted into $\vert\phi\rangle\otimes\vert\chi\rangle$ with
complete success. In such a transformation, the entanglement of
the catalyst is not consumed, although it takes part in the
transformation. Catalysis is also useful in almost all conversion
processes where it improves upon the conversion probability.

Expressing in terms of the Schmidt coefficients and considering
only the cases where catalysis helps achieve unit probability of
success, we basically have situations where $x$ is not majorized
by $y$, but there is a sequence $c$ such that $x\otimes c$ is
majorized by $y\otimes c$. Following Nielsen, for two sequences of
non-negative numbers $x$ and $y$ with $n$-elements, we will say
that $x$ is \emph{trumped} by $y$ (written $x\prec_T y$) if there
exists another sequence $c$ such that $x\otimes c\prec y\otimes
c$. It is easy to see that in such cases the catalyst sequence $c$
can be chosen from strictly positive numbers.

A lot of research has been directed to understand the catalytic
transformations\cite{Feng1} and to analyze the mathematical
structure of the trumping
relation.\cite{Daftuar,Feng2,Duan1,Feng3} One of the open problems
is to find a way to decide if two given sequences $x$ and $y$
satisfies the trumping relation. The purpose of this article is to
give all necessary and sufficient conditions for this relation.
This problem has been partially solved by Aubrun and
Nechita\cite{Aubrun1,Aubrun2}, who work with stochastic tools to
describe the closure of a set constructed with the trumping
relation. The methods used in this article, however, are quite
similar to those of a recent study that provided an expression for
the catalytic conversion probability.\cite{SuperTrump} But, as the
trumping relation necessarily implies that the two sequences have
the same sum, the mathematical details of the proofs given below
are more complicated than the ones in that article.

It appears that the necessary and sufficient conditions for the
trumping relation can be expressed in terms of the power means,
which are defined as
\begin{equation}
  A_\nu(x)=\left(\frac{\sum_{i=1}^n   x_i^\nu}{n}\right)^\frac{1}{\nu}\quad,
\end{equation}
and the entropy of entanglement,
\begin{equation}
  \sigma(x)=-\sum_{i=1}^n x_i\ln x_i \quad.
\end{equation}
The main theorem that we would like to prove is the following.

\textbf{Theorem 1.} For two $n$-element sequences of non-negative
numbers $x$ and $y$ such that $x$ has non-zero elements and the
sequences are distinct (i.e., $x^\uparrow\neq y^\uparrow$), the
relation $x\prec_T y$ is equivalent to the following three
inequalities
\begin{eqnarray}
  A_\nu(x) &>& A_\nu(y)\quad,\quad\forall~\nu\in(-\infty,1)\quad,\label{cond:A}  \\
  A_\nu(x) &<& A_\nu(y)\quad,\quad\forall~\nu\in(1,+\infty)\quad,\label{cond:B}  \\
  \sigma(x) &>& \sigma(y)\quad,  \label{cond:C}
\end{eqnarray}
where all inequalities are strict.

Note that by the continuity of the power mean function against
$\nu$, the requirement that the sequences $x$ and $y$ have the
same sum is included in the conditions (\ref{cond:A}) and
(\ref{cond:B}). Moreover, the limits of these inequalities at
$\nu=-\infty$ and $\nu=+\infty$ imply that the minimum and maximum
elements of the sequences satisfy the respective inequalities
$x_1^\uparrow\ge y_1^\uparrow$ and $x_n^\uparrow\le y_n^\uparrow$,
but these do not have to be strict. Note also that all of these
three conditions can be equivalently expressed as the strict
positivity of the function
\begin{equation}
  R_\nu=\frac{1}{\nu-1}\ln\frac{A_\nu(y)}{A_\nu(x)}
\end{equation}
for all finite values of $\nu$, where the value $R_1$ corresponds
to the difference of the entropies.

Notice that some of the results in Ref. \onlinecite{FengMLOCC}
about the connection between the multiple-copy entanglement
transformation, a related phenomenon discovered by Badyopadhyay
\textit{et al.}\cite{Bandyopadhyay}, and the trumping relation can
be easily understood in view of the conditions
(\ref{cond:A}-\ref{cond:C}). For any sequence $x$ and any integer
$k>1$, we have $A_\nu(x^{\otimes k})=A_\nu(x)^k$ and
$\sigma(x^{\otimes k})=k\sigma(x)$. As a result, if $k$ copies of
a state with coefficients $x$ can be transformed into $k$ copies
of another state with coefficients $y$, either with or without
catalysis, then $x$ must be trumped by $y$.

The article is organized as follows: In section \ref{sec:sec1}, a
few relations related to majorization are given and a key lemma is
proved. Then, in section \ref{sec:proof}, the theorem is proved.
Finally, section \ref{sec:conclusions} contains the conclusions.

\section{A few Relations and a Key Lemma}
\label{sec:sec1}

The following facts about the majorization and the trumping
relation will be used occasionally.
\begin{itemize}
\item[(1)] If $x\prec y$, then for any convex function $f$, we have
\begin{equation}
  \sum_{i=1}^n f(x_i) \le  \sum_{i=1}^n f(y_i)\quad.
  \label{eq:convexineq}
\end{equation}
Moreover, if $x^\uparrow\neq y^\uparrow$ and $f$ is strictly
convex, then the inequality above is strict.

\item[(2)] For any sequence $x$, we define the
\emph{characteristic function} $H_x(t) = \sum_{i=1}^n (t-x_i)^+$
where $(\alpha)^+=\max(\alpha,0)$ denotes the positive-part
function. For non-negative sequences $x$ and $y$ which have the
same sum ($\sum x_i=\sum y_i$), the relation $x\prec y$ can be
equivalently stated as
\begin{equation}
H_x(t) \le H_y(t)\quad\forall~t\ge0\quad.
\end{equation}

\item[(3)]  For the cross-product of two sequences we have
$H_{x\otimes c} =\sum_\ell c_\ell H_x(t/c_\ell)$.

\item[(4)] Relation $\prec$ and $\prec_T$ are partial orders on
all $n$-element sequences (up to equivalence under rearrangement).
Moreover, $x\prec y$ implies $x\prec_T y$.

\item[(5)] Let $z$ be any sequence of strictly positive numbers
and let $x\oplus z$ denote the sequence obtained by concatenating
the elements of $z$ to those of $x$. Then $x\prec_T y$ iff
$x\oplus z\prec_T y\oplus z$. Moreover, $x$ and $y$ satisfies the
inequalities (\ref{cond:A}-\ref{cond:C}) iff $x\oplus z$ and
$y\oplus z$ satisfies the same inequalities. As a result, nothing
will be lost from generality if a proof of the Theorem 1 is given
for sequences that have no common elements.
\end{itemize}

The proof of the sufficiency of the conditions
(\ref{cond:A}-\ref{cond:C}) is based on the following key lemma.

\textbf{Lemma:} If a polynomial $\gamma(s)$ has no positive roots
and $\gamma(0)>0$ then
\begin{itemize}
\item[(a)] it can be expressed as $\gamma(s)=b(s)/a(s)$ where
$a(s)$ and $b(s)$ are polynomials with non-negative coefficients.

\item[(b)] Moreover, $a(s)$ can be chosen as a polynomial with
integer coefficients.
\end{itemize}

\textit{Proof:} The lemma can easily be generalized to polynomials
which have a root at $s=0$, but for our purposes the above form is
sufficient. For part (a), we first provide the proof for a second
degree polynomial with complex roots, e.g., $\gamma(s)=1-2\xi
s+\lambda s^2$ where $\lambda>\xi^2$. For $\xi\le0$, there is
nothing to be proved as $\gamma$ has already non-negative
coefficients. For $\xi>0$, $\gamma(s)$ can be written as the ratio
$b(s)/a(s)$ where
\begin{eqnarray}
a(s) &=& \sum_{k=0}^{2N-1} (1+\lambda s^2)^k (2\xi s)^{2N-1-k}\quad,\\
b(s) &=& (1+\lambda s^2)^{2N}-(2\xi s)^{2N}\quad,
\end{eqnarray}
and $N$ is a sufficiently large integer so that the inequality
\begin{equation}
  \frac{1}{4} \left(\frac{(2N)!}{N!^2}\right)^\frac{1}{N} \ge
  \frac{\xi^2}{\lambda}\quad,
\end{equation}
is satisfied. It is always possible to find such an $N$ since the
left-hand side has a limit $1$ when $N$ goes to infinity and the
right-hand side is strictly less than $1$. For such a choice of
$N$, both $a(s)$ and $b(s)$ will have non-negative coefficients.

For a general polynomial $\gamma$ which has no positive root, we
first express it as a product of its irreducible factors as
\begin{equation}
  \gamma(s)= A \prod_i (1+\zeta_i s)\prod_i(1-2\xi_i s+(\xi_i^{2}+\eta_i^{2})s^2)\quad,
\end{equation}
where $A>0$, $-1/\zeta_i$ are the real roots of $\gamma$
(therefore $\zeta_i>0$) and $(\xi_i\pm i\eta_i)^{-1}$ are the
complex roots of $\gamma$. As the quadratic factors can be
expressed as a ratio and the rest is simply a polynomial with
non-negative coefficients, $\gamma$ can be expressed as a ratio of
two polynomials with non-negative coefficients. Note that, as
$\gamma$ has no root at $0$, both $a(s)$ and $b(s)$ can be chosen
as polynomials having a non-zero constant term.

Before passing on to the proof of the statement (b), we first show
that the polynomial $b(s)$ can always be chosen such that all of
its coefficients are strictly positive. For this, consider a
degree $m$ solution for $b(s)$, i.e., $b(s)=\sum_{k=0}^m b_k s^k$
where $b_0>0$, $b_m>0$ and $b_k\ge0$ for all $1\le k<m$. Let
$e(s)=1+s+\cdots+s^{m-1}$. Then $e(s)b(s)$ is a polynomial with
degree $2m-1$ and all of its $2m$ coefficients are positive.
Moreover, the polynomials $e(s)b(s)$ and $e(s)a(s)$ satisfy the
conditions of part (a). Therefore, $b(s)$ can be chosen to have
non-zero coefficients.

For the proof of part (b), suppose that $b(s)$ is a degree $m$
polynomial with positive coefficients and let $\beta=\min_{0\le
k\le m} b_k$ be the minimum of those. Let
\begin{equation}
  \epsilon=\frac{\beta}{\sum_k\vert\gamma_k\vert}\quad,
\end{equation}
where $\gamma_n$ are the coefficients of the polynomial
$\gamma(s)$. Define a new polynomial $\bar{a}(s)$ such that it has
the same degree as $a(s)$ and its coefficients are chosen from
rational numbers such that
\begin{equation}
  \vert \bar{a}_k-a_k\vert\le\epsilon\quad k=0,1,\ldots,N\quad,
\end{equation}
where $\bar{a}_k$ and $a_k$ are the coefficients of $\bar{a}(s)$
and $a(s)$ respectively. As the rational numbers are dense, this
can always be done. If $\bar{a}(s)\gamma(s)=\bar{b}(s)$, then the
coefficients $\bar{b}(s)$ satisfy
\begin{eqnarray}
  \bar{b}_k-b_k &=& \sum_{\ell}
  (\bar{a}_\ell-a_\ell)\gamma_{k-\ell}
   \ge  -\epsilon\sum_\ell \vert\gamma_\ell\vert \ge   -\beta\quad.
\end{eqnarray}
Therefore, $\bar{b}_k\ge b_k-\beta\ge0$, i.e., $\bar{b}(s)$ has
non-negative coefficients as desired. Multiplying $\bar{a}(s)$ by
the common denominator of its coefficients gives a polynomial with
integer coefficients.$\Box$

\section{Proof of Theorem 1}
\label{sec:proof}

Proof of the necessity of the conditions
(\ref{cond:A}-\ref{cond:C}) for the trumping relation is trivial.
Given that there is a catalyst $c$ so that we have $x\otimes
c\prec y\otimes c$, we use the strict inequality of
(\ref{eq:convexineq}) for the following strictly convex functions:
$f(t)=t^\nu$ for $\nu>1$ and $\nu<0$, $f(t)=-t^\nu$ for $0<t<1$,
$f(t)=-\ln t$ and $f(t)=t\ln t$. All inequalities
(\ref{cond:A}-\ref{cond:C}) follow from these.

The proof of the sufficiency of the conditions
(\ref{cond:A}-\ref{cond:C}) is lengthy and requires us to separate
it into three special cases. The key proof is for case A, where
only the sequences which can be expressed as integer powers of a
common number is considered. Case B concentrates on non-zero
sequences and uses the stability of the sufficiency conditions
under small changes of sequences to reduce the problem to case A.
Finally, case C deals with the situation where $y$ has zero
elements.

\textbf{Case A.} $y$ has strictly positive elements such that
$y_i=K\omega^{\alpha_i}$ and $x_i=K\omega^{\beta_i}$ for some
integers $\alpha_i$ and $\beta_i$ and for some numbers $K>0$ and
$\omega>1$.

\textit{Proof:} Without loss of generality, it is assumed that $x$
and $y$ have no common elements and they are arranged in
non-decreasing order. The smallest of the exponents is $\alpha_1$
which can be set equal to 0 by a redefinition of $K$. Finally,
both $x$ and $y$ can be divided by $K$ which is equivalent to
setting $K=1$. As a result, it is not required that the sequences
are normalized (i.e., they do not add up to 1). Since
$\alpha_1=0$, all other exponents satisfy $\alpha_i\ge0$ and
$\beta_i>0$.

Let the polynomial $\Gamma(s)$ be defined as
\begin{equation}
  \Gamma(s)=\sum_{i=1}^n (s^{\alpha_i}-s^{\beta_i})=\sum_{k}\Gamma_k  s^k\quad.
\end{equation}
First, note that $\Gamma(s)$ has simple roots at $s=1$ and
$s=\omega$. This can be simply seen by evaluating its derivative
at these points,
\begin{eqnarray}
  \Gamma^\prime(1) &=& \sum_{i=1}^n (\alpha_i -\beta_i) <0\quad,\\
  \Gamma^\prime(\omega)  &=&\frac{\sigma(x)-\sigma(y)}{\ln\omega}>0\quad,
\end{eqnarray}
where the former strict inequality follows from (\ref{cond:A}) at
$\nu=0$ and the latter follows from (\ref{cond:C}). The fact that
$x$ and $y$ are not normalized do not invalidate the latter
inequality.

Therefore, $\gamma(s)=\Gamma(s)/((1-s)(1-s/\omega))$ is a
polynomial. It can be seen that $\gamma(0)>0$. Moreover, we will
show that $\gamma(s)$ has no positive root. For this purpose let
$s=\omega^\nu$ where $\nu$ is any real number ($\nu=0$ and $\nu=1$
can be excluded if desired). Then
\begin{eqnarray}
  \gamma(\omega^\nu) = \frac{1}{(1-\omega^\nu)(1-\omega^{\nu-1})}
  \sum_{i=1}^n(y_i^\nu-x_i^\nu)\quad,
\end{eqnarray}
which can be seen to be strictly positive by virtue of
(\ref{cond:A}) and (\ref{cond:B}) for all values of $\nu$. (For
$\nu=0$ and $\nu=1$, we have seen above that $\gamma(s)$ has no
root at $1$ and $\omega$).

By the lemma, there are two polynomials $a(s)$ and $b(s)$ with
non-negative coefficients such that $a(s)\gamma(s)=b(s)$ and
$a(s)$ has integral coefficients. The constant coefficients $a(0)$
and $b(0)$ will also be chosen to be non-zero. In terms of
$\Gamma$, the relation can be expressed as
\begin{equation}
  a(s)\Gamma(s)=(1-s)(1-s/\omega) b(s)\quad.
\end{equation}

Let $a(s)$ has degree $N$. The catalyst sequence $c$ will be
chosen from the numbers $\omega^k$ which are repeated $a_k$ times
($k=0,1,\ldots,N$). In that case, the characteristic function of
$c$ is
\begin{equation}
  H_c(t)=\sum_{k=0}^N a_k (t-\omega^k)^+\quad.
\end{equation}
We would like to show that the function
\begin{eqnarray}
\Delta(t) &=& H_{y\otimes c}(t)- H_{x\otimes c}(t) \\
        &=& \sum_{i=1}^n y_i H_c(t/y_i)- x_i H_c(t/x_i) \\
        &=& \sum_\ell \Gamma_\ell \omega^\ell H_c(t\omega^{-\ell})\\
        &=& \sum_{k,\ell} a_k \Gamma_\ell (t-\omega^{k+\ell})^+\quad,
\end{eqnarray}
is non-negative for all $t\ge0$. First note that
\begin{equation}
  \Delta(t)=\sum_{m=0}^{M+1} (f_m-f_{m-1})(t-\omega^m)^+
\end{equation}
where $f(s)=(1-s/\omega)b(s)$, $f_m$ are coefficients of the
polynomial $f(s)$ and we have chosen $f_{-1}=0$ for simplicity.
Here $M$ is the degree of $f$ ($M+1$ is the degree of
$a(s)\Gamma(s)$). Since $\Delta(t)$ is a piecewise linear
function, for showing its positivity, it is sufficient to look at
its value at the turning points and at the $0$ and $\infty$
limits. First note that $\Delta(t)=0$ for $t\le1$ and $\Delta(t)$
is constant for $t\ge \omega^{M+1}$. As a result, we only need to
check the values of $\Delta(t)$ at
$t=\omega,\omega^2,\ldots,\omega^{M+1}$. For any $1\le k\le M+1$,
\begin{eqnarray}
  \Delta(\omega^k) &=&\sum_{m=0}^{k-1}(f_m-f_{m-1})(\omega^k  -  \omega^m)\\
%    &=&(\omega-1) \sum_{m=0}^{k-1}(f_m-f_{m-1})\sum_{\ell=0}^{k-m-1}\omega^{m+\ell}\\
    &=&(\omega-1) \sum_{m=0}^{k-1}(f_m-f_{m-1})\sum_{p=m}^{k-1}\omega^p\\
    &=&(\omega-1) \sum_{p=0}^{k-1}\omega^p \sum_{m=0}^p(f_m-f_{m-1})\\
    &=& (\omega-1) \sum_{p=0}^{k-1}f_p \omega^p \quad.
\end{eqnarray}
Finally, as $f(s)=(1-s/\omega)b(s)$, the coefficients of these
polynomials satisfy
\begin{equation}
  f_p = b_p - \frac{b_{p-1}}{\omega}\quad,
\end{equation}
where $b_{-1}=0$, which leads to
\begin{equation}
 \Delta(\omega^k) = (\omega-1)b_{k-1} \omega^{k-1} \ge 0\quad.
\end{equation}
This completes the proof of $\Delta(t)\ge0$ for all $t\ge0$. It
also shows that $x\otimes c\prec y\otimes c$. Therefore, $x\prec_T
y$.$\Box$

Before passing on to the next case, we first state another theorem
that shows the stability of the inequalities
(\ref{cond:A}-\ref{cond:C}) against small variations in sequences
$x$ and $y$. Since only sequences with non-zero elements will be
considered in the next case, the distance between two sequences
will be measured by the deviation of the ratio of the
corresponding elements from $1$. For two sequences $x$ and
$\bar{x}$ which has no zero elements, the distance between them is
defined as
\begin{equation}
  D(x;\bar{x})=\max_{i}  \left\vert\ln\frac{x_i}{\bar{x}_i}\right\vert\quad.
\end{equation}
The following theorem expresses the stability of the conditions
(\ref{cond:A}-\ref{cond:C}).

\textbf{Theorem 2.} Let $x$ and $y$ be $n$-element sequences
formed from positive numbers such that $x_1^\uparrow>y_1^\uparrow$
and $x_n^\uparrow<y_n^\uparrow$. If $x$ and $y$ satisfy the
inequalities (\ref{cond:A}), (\ref{cond:B}) and (\ref{cond:C}),
then there is a positive number $\epsilon$ such that whenever
$D(x;\bar{x})\le\epsilon$ and $D(y;\bar{y})\le\epsilon$, and $\sum
\bar{x}_i=\sum \bar{y}_i=\sum x_i$, the sequences $\bar{x}$ and
$\bar{y}$ satisfy the same strict inequalities.

The proof of Theorem 2 is postponed to Appendix \ref{app:A}. This
result will be used in the proof of the next case.

\textbf{Case B.} $y$ has strictly positive elements.

The proof will be carried out by choosing two new sequences
$\bar{x}$ and $\bar{y}$ which are sufficiently near to $x$ and $y$
such that Theorem 2 can be invoked, and it will be made sure that
$\bar{x}$ and $\bar{y}$ satisfy the conditions considered in case
A. Without loss of generality, it is assumed that $x$ and $y$ are
normalized ($\sum x_i=\sum y_i=1$) and they are arranged in
non-decreasing order. Let $H=\sigma(x)-\sigma(y)>0$ be the entropy
difference of these sequences and let $L=\vert\ln
y_1^\uparrow\vert$. Note that the logarithm of all elements are
bounded by $L$, i.e., $\vert\ln y_i\vert\le L$ and $\vert\ln
x_i\vert\le L$. Let $\epsilon_0$ be a positive number such that
whenever $D(x,\bar{x})\le\epsilon_0$ and
$D(y,\bar{y})\le\epsilon_0$, the sequences $\bar{x}$ and $\bar{y}$
satisfy all the inequalities in (\ref{cond:A}-\ref{cond:C}). The
positive number $\epsilon$ is chosen such that
\begin{equation}
  \epsilon <  \min\left(\frac{\epsilon_0}{2},\frac{1}{8n},\frac{1}{n^2},\frac{H}{96nL}\right)\quad.
\label{eq:epsilon_definition}
\end{equation}
First note that the definition above implies that $\epsilon<L$, an
inequality that will be used below.

We will define $\alpha_i$ and $\beta_i$ to be some rational
approximations to numbers $\ln y_i$ and $\ln x_i$ respectively.
Let $\phi_i$ and $\theta_i$ represent the deviation of these
rational approximations from the true values,
\begin{eqnarray}
  \alpha_i &=& \ln y_i + \phi_i\quad,\\
  \beta_i  &=& \ln x_i + \theta_i\quad.
\end{eqnarray}
As the rational numbers are dense, these deviations can be chosen
essentially arbitrarily. But, for our purposes, we are going to
choose them as
\begin{eqnarray}
  \frac{\epsilon}{2n} \le \phi_i &\le& \frac{\epsilon}{n}\quad\textrm{for}~~1\le i\le n-1\quad,\label{eq:phicond1} \\
  \left\vert \sum_{i=1}^n y_i \phi_i \right\vert &\le&  \epsilon^2\quad.\label{eq:phicond2}
\end{eqnarray}
In other words, the rational approximations $\alpha_i$ for all
elements excepting the last one are to be chosen such that the
corresponding deviations $\phi_i$ are positive and small, but they
are also required to be sufficiently far away from zero. The last
element is an exception. In that case $\alpha_n$ has to be chosen
as a rational number so that this time the sum in
(\ref{eq:phicond2}) is made very small. In that case, $\phi_n$
does not need to be positive. Note that the conditions
(\ref{eq:phicond1}) and (\ref{eq:phicond2}) provides $n$ separate
intervals to choose $\alpha_i$ from. As rational numbers are
dense, all of $\alpha_i$ can be chosen as rational numbers.
Similarly, we define $\beta_i$ and the corresponding deviations
$\theta_i$ such that
\begin{eqnarray}
  -\frac{\epsilon}{n}\le \theta_i &\le& -\frac{\epsilon}{2n}   \quad\textrm{for}~~1\le i\le n-1\quad,\label{eq:phicond3} \\
  \left\vert \sum_{i=1}^n x_i \theta_i \right\vert &\le&  \epsilon^2\quad,\label{eq:phicond4}
\end{eqnarray}
where the deviations for the first $n-1$ elements are chosen this
time to be negative. Similar comments apply in here.

Below, however, we will need a uniform bound on all the
deviations. For this purpose, note the following bound on $\phi_n$
\begin{eqnarray}
   y_n \vert\phi_n\vert &\le& \epsilon^2 +\sum_{i=1}^{n-1} y_i \vert \phi_i\vert \le \epsilon^2   +(1-y_n)\frac{\epsilon}{n}\\
   \vert\phi_n\vert &\le& \frac{\epsilon^2}{y_n} +\left(\frac{1}{y_n}-1\right)\frac{\epsilon}{n}\\
        &\le& n\epsilon^2 +(n-1)\frac{\epsilon}{n}\le \epsilon
\end{eqnarray}
where we have used the fact that $y_n\ge1/n$ for the maximum element of $y$.
Therefore, the following uniform bounds can be placed on all deviations
\begin{equation}
  \vert\phi_i\vert \le \epsilon\quad,\quad \vert\theta_i\vert \le \epsilon \quad\textrm{for}~i=1,2,\ldots,n\quad,
\end{equation}
where the bounds on $\theta_i$ follow by a similar analysis. For
most of the following, we will use these uniform bounds. The
stricter bounds given in (\ref{eq:phicond1}) and
(\ref{eq:phicond3}) will only be necessary at the very end. The
following bounds on the rational approximations will be
occasionally used: $\vert \alpha_i\vert \le\vert \ln
y_i\vert+\vert\phi_i\vert\le L+\epsilon\le 2L$ and similarly
$\vert \beta_i\vert\le 2L$.

Consider the following function
\begin{equation}
  F(\lambda)=\sum_{i=1}^n\left(e^{\lambda \alpha_i}-e^{\lambda \beta_i}\right)\quad.
\end{equation}
Our first job is to establish that this function has a root near
1, i.e., there is a number $\lambda_0$, which is very close to $1$
such that $F(\lambda_0)=0$. Once this problem is solved, the two
new sequences $\bar{x}$ and $\bar{y}$ can be defined as
\begin{eqnarray}
  \bar{x}_i &=& \frac{e^{\lambda_0\beta_i}}{Z_0}\quad,\\
  \bar{y}_i &=& \frac{e^{\lambda_0\alpha_i}}{Z_0}\quad,
\end{eqnarray}
where $Z_0 = \sum_{i=1}^n e^{\lambda_0 \alpha_i} = \sum_{i=1}^n
e^{\lambda_0 \beta_i}$. In that case, both $\bar{x}$ and $\bar{y}$
are normalized sequences. However, in order to reach to the final
conclusion, we also need to place bounds on the deviation of both
$\lambda_0$ and $Z_0$ from $1$. Therefore, the following analysis
of bounds is needed.

First, we must show that $F(\lambda)$ has a root somewhere near
$1$. For this purpose, we look at the value of $F(1)$. By using
the following inequalities satisfied by the exponential function,
$1+t\le e^t\le 1+t+t^2$ for all $\vert t \vert\le1$, the following
bounds can be placed on the first term of $F(1)$,
\begin{eqnarray}
\sum_{i=1}^n e^{\alpha_i} &=& \sum_{i=1}^n y_i e^{\phi_i} \\
 &\ge& \sum_{i=1}^n y_i (1+\phi_i)\ge 1-\epsilon^2\quad,\label{eq:alpha-sum-bound1}\\
\sum_{i=1}^n e^{\alpha_i}& \le& \sum_{i=1}^n y_i
(1+\phi_i+\phi_i^2) \le 1+2\epsilon^2 \quad.\label{eq:alpha-sum-bound2}
\end{eqnarray}
Same bounds can also be placed for the second term as well, which
lead to
\begin{equation}
  \vert F(1)\vert \le 3\epsilon^2\quad,
\end{equation}
a very small quantity, which indicates that a root is very close
to $1$.

However, to verify that there is root around 1 and to place a
bound on the deviation of the root from 1, we must make sure that
the derivative $F^\prime(\lambda)$ does not rapidly go to zero
around $\lambda=1$. For this purpose, a lower bound will be placed
on the derivative for $\vert\lambda-1\vert\le \epsilon/L$. First,
note that
\begin{eqnarray}
\sum_{i=1}^n \alpha_i e^{\lambda \alpha_i} &=&-\sigma(y)+\sum_{i=1}^n y_i\phi_i  \\
    & & + \sum_{i=1}^n y_i \alpha_i \left(e^{(\lambda-1)\ln y_i+\lambda \phi_i}-1\right)\quad,
\end{eqnarray}
and the argument of the exponential is small as
\begin{equation}
\left\vert(\lambda-1)\ln y_i+\lambda \phi_i\right\vert \le
\frac{\epsilon}{L} L +\left(1+\frac{\epsilon}{L}\right)\epsilon
\le 3\epsilon\quad.
\end{equation}
Now, using $\vert e^t-1\vert\le \vert t\vert+t^2\le 2\vert t\vert$ for all $\vert t\vert\le1$, we can
find the following lower bound on the expression above
\begin{eqnarray}
\sum_{i=1}^n \alpha_i e^{\lambda \alpha_i} &\ge &-\sigma(y)-\epsilon^2 -2L \cdot 6\epsilon\\
    &\ge& -\sigma(y)-13L\epsilon
\end{eqnarray}
Similar analysis for the second term of $F^\prime(\lambda)$ gives
\begin{eqnarray}
\sum_{i=1}^n  \beta_i e^{\lambda \beta_i}
&\le&-\sigma(x)-13L\epsilon \quad.
\end{eqnarray}
Both of these give the following lower bound on the derivative $F^\prime(\lambda)$
for $\vert\lambda-1\vert\le\epsilon/L$,
\begin{equation}
  F^\prime(\lambda)\ge H -26 L\epsilon >   \frac{1}{2}H\quad.
\end{equation}

By using the lower bound given above it is possible to see that
$F(1+\epsilon/L)$ is positive and $F(1-\epsilon/L)$ is negative.
This guarantees the presence of the root in the specified
interval. But, this interval is too large for our purposes, and we
need to find a better bound on the place of the root. Using
$F(\lambda_0)=0$, we can get
\begin{eqnarray}
  -F(1) &=& \int_1^{\lambda_0} F^\prime(\lambda)d\lambda \quad,\\
  \vert F(1)\vert &\ge& \vert\lambda_0-1\vert \frac{H}{2}\quad,\\
  \vert\lambda_0-1\vert &\le& \frac{2\vert F(1)\vert }{H}\le \frac{6\epsilon^2}{H}\quad.
\end{eqnarray}
In other words, the root is very close to the value 1.

One final bound, this time a bound on $\ln Z_0$ will be needed.
For this, we first note that
\begin{equation}
Z_0=\sum_{i=1}^n e^{\alpha_i}e^{(\lambda_0-1) \alpha_i}\le
\left(\sum_{i=1}^n e^{\alpha_i}\right)e^{+2\vert\lambda_0-1\vert
L}\quad,
\end{equation}
and a similar analysis for the lower bound gives
\begin{equation}
\left\vert \ln Z_0\right\vert\le \left\vert\ln\left(\sum_{i=1}^n
e^{\alpha_i}\right)\right\vert+2\vert\lambda_0-1\vert L\quad.
\end{equation}
Finally, (\ref{eq:alpha-sum-bound1}) and (\ref{eq:alpha-sum-bound2}) gives
\begin{equation}
\left\vert \ln\left(\sum_{i=1}^n e^{\alpha_i}\right)\right\vert\le 2\epsilon^2
\end{equation}
where we have used the fact that $t-1\ge\ln t\ge (t-1)/t$.
As a result, we get
\begin{equation}
  \vert\ln Z_0\vert \le \left(2 + \frac{12L}{H}\right)\epsilon^2\quad.
\end{equation}

Now, it is possible to show that the sequences $\bar{x}$ and
$\bar{y}$ satisfy all the required properties to complete the
proof. First, we will show that $x$ is majorized by $\bar{x}$. For
this reason, we will look at the ratio $x_i/\bar{x}_i$ for
$i=1,2,\ldots,n-1$, i.e., for all elements except the last one.
Here, we will make use of the upper bounds given in
(\ref{eq:phicond3}) as
\begin{eqnarray}
\ln\frac{x_i}{\bar{x}_i}&=& -\theta_i+(1-\lambda_0)\beta_i+\ln Z_0 \\
&\ge& \frac{\epsilon}{2n} -\left(2 +
\frac{24L}{H}\right)\epsilon^2 \ge0\quad,
\end{eqnarray}
where the last inequality can be obtained simply by inspecting
(\ref{eq:epsilon_definition}). In other words, we have
$x_i\ge\bar{x}_i$ for all $i<n$. The conclusion $x\prec\bar{x}$
then follows. By the same method, it can be shown that $\bar{y}$
is majorized by $y$ as
\begin{eqnarray}
\ln\frac{\bar{y}_i}{y_i} &=& \phi_i+(\lambda_0-1)\alpha_i-\ln Z_0\\
  &\ge& \frac{\epsilon}{2n} -\left(2 +\frac{24L}{H}\right)\epsilon^2 \ge0\quad,
\end{eqnarray}
in other words $\bar{y}_i\ge y_i$ for all $i<n$ and therefore
$x\prec_T y$.

Finally, we have
\begin{eqnarray}
  D(x;\bar{x}) &=& \max_i \vert \theta_i-(1-\lambda_0)\beta_i-\ln Z_0\vert \\
    &\le & \epsilon+ \left(2 +\frac{24L}{H}\right)\epsilon^2< \epsilon_0\quad,
\end{eqnarray}
and similarly   $D(y;\bar{y})<\epsilon_0$. Therefore, the
inequalities (\ref{cond:A}-\ref{cond:C}) are also satisfied by
$\bar{x}$ and $\bar{y}$. It is easy to see that $\bar{x}$ and
$\bar{y}$ satisfy the conditions of case A. The number $\omega$ is
given as $\exp(\lambda_0/\mathcal{N})$ where $\mathcal{N}$ is the
common denominator of the rational numbers $\alpha_i$ and
$\beta_i$. As a result, the conclusion $\bar{x}\prec_T\bar{y}$
follows. Combined with $x\prec\bar{x}$ and $\bar{y}\prec y$, it
leads to the desired result $x\prec_T y$.$\Box$

\textbf{Case C.} $y$ has zero components.

Without loss of generality, it is supposed that $x$ and $y$ are
normalized, they are arranged in non-decreasing order and have no
common elements. Let $y$ have $m$ zeros, i.e.,
$y_1=y_2=\cdots=y_m=0$ and $0<y_{m+1}\le\cdots\le y_n$. Let
$z^\epsilon$ be a sequence defined as follows,
\begin{eqnarray}
  z^\epsilon_i &=& \epsilon\quad\textrm{for}~i=1,2,\ldots,m\quad,\\
  z^\epsilon_i &=& (1-m\epsilon)y_i \quad\textrm{for}~i=m+1,\ldots,n\quad.
\end{eqnarray}
where $\epsilon$ is a non-negative parameter. We will only be
interested in the values of $\epsilon$ in the range
$\epsilon\le(y_{m+1}^{-1}+m)^{-1}$, where $z^\epsilon$ is arranged
in increasing order. It is easy to see that all such sequences are
related to each other by the majorization relation, i.e., if
$\epsilon_A>\epsilon_B$ then $z^{\epsilon_A}\prec z^{\epsilon_B}$.
As $z^0=y$, we have $z^\epsilon\prec y$ for all values of
$\epsilon$ in the range considered.

Our job is to show that if $\epsilon$ is sufficiently small, then
$x$ and $z^\epsilon$ satisfy the inequalities
(\ref{cond:A}-\ref{cond:C}). This is a straightforward but
laborious procedure which is detailed below. For this purpose,
different intervals of $\nu$ values will be considered separately
and for each interval, the existence of a separate upper bound for
$\epsilon$ will be provided.

(a) \textit{For $\nu\le0$:} The quantity $\epsilon_1=y_n
(x_1/y_n)^{n/m}$ is a possible upper bound for this range. Let
$\epsilon<\epsilon_1$. For the special case $\nu=0$, we have
\begin{eqnarray}
\frac{A_0(x)}{A_0(z^\epsilon)} &=& \left(\frac{\prod_{i=1}^n
x_i}{\epsilon^m (1-m\epsilon)^{n-m}\prod_{i=m+1}^n y_i}
\right)^\frac{1}{n} \\
    &\ge &
    \frac{x_1}{y_n}\left(\frac{y_n}{\epsilon}\right)^\frac{m}{n}>1\quad.
\end{eqnarray}
For all negative values of $\nu$ we make use of Bernoulli's
inequality, which states that $\alpha^r-1\ge r(\alpha-1)$ for any
$r\ge1$ and any positive number $\alpha$, to reach
\begin{equation}
  m(\epsilon^\nu-y_n^\nu) > n(x_1^\nu-y_n^\nu)\quad.
\end{equation}
This then leads to
\begin{eqnarray}
\sum_{i=1}^n (z^\epsilon_i)^\nu &=& m\epsilon^\nu + (1-m\epsilon)^\nu\sum_{i=m+1}^n y_i^\nu \\
    &>& m\epsilon^\nu + (n-m)y_n^\nu > nx_1^\nu\ge \sum_{i=1}^n x_i^\nu \quad.
\end{eqnarray}
As a result, we conclude that $A_\nu(x)>A_\nu(z^\epsilon)$ for all
$\nu\le0$ whenever $\epsilon<\epsilon_1$.

(b) \textit{For $0<\nu\le1/2$:} The function
\begin{equation}
  J_\nu = \left(\frac{\sum_{i=1}^n x_i^\nu-\sum_{i=m+1}^n
  y_i^\nu}{m}\right)^\frac{1}{\nu}\quad
\end{equation}
is strictly positive in the interval $(0,1/2]$ and moreover it has
a strictly positive limit at $\nu=0$. Therefore,
$\epsilon_2=\min_{\nu\in[0,1/2]} J_\nu$ is a positive number. If
$\epsilon<\epsilon_2$, we have
\begin{eqnarray}
  \sum_{i=1}^n x_i^\nu> m\epsilon^\nu+\sum_{i=m+1}^n y_i^\nu>\sum_{i=1}^n
  (z^\epsilon_i)^\nu\quad,
\end{eqnarray}
which leads to $A_\nu(x)>A_\nu(z^\epsilon)$ in this interval.

(c) \textit{For $2\le\nu$:} Let $K$ be defined as
\begin{equation}
  K = \max_{\nu\in[2,\infty]} \frac{A_\nu(x)}{A_\nu(y)}\quad,
\end{equation}
which is a positive number such that $K<1$. Note that, as $x$ and
$y$ have no common elements, the ratio above at $\nu=+\infty$
gives $x_n^\uparrow/y_n^\uparrow$ which is smaller than $1$. Let
$\epsilon_3=(1-K)/m$. Then, for any $\epsilon<\epsilon_3$ and for
all $\nu\ge 2$ we have
\begin{eqnarray}
  \sum_{i=1}^n (z^\epsilon_i)^\nu &>& (1-m\epsilon)^\nu \sum_{i=m+1}^n  y_i^\nu \\
    &>& K^\nu \sum_{i=m+1}^n  y_i^\nu >\sum_{i=1}^n x_i^\nu\quad.
\end{eqnarray}
This shows the desired inequality, $A_\nu(z^\epsilon)>A_\nu(x)$.

(d) \textit{For $1/2\le \nu \le 2$:} Let
\begin{equation}
R_\nu=\frac{1}{\nu-1}\ln \frac{A_\nu(y)}{A_\nu(x)}\quad.
\end{equation}
The inequalities (\ref{cond:A}-\ref{cond:C}) imply that $R_\nu$ is
a strictly positive continuous function in the interval
considered. Therefore, the minimum $M=\min_{\nu\in[1/2,2]} R_\nu$
is a positive number. Let
\begin{equation}
R_\nu(\epsilon)=\frac{1}{\nu-1}\ln\frac{A_\nu(z^\epsilon)}{A_\nu(x)}
\end{equation}
Since all sequences $z^\epsilon$ are related into each other by
the majorization relation, for any $\epsilon_A>\epsilon_B$ we have
$R_\nu(\epsilon_A)\le R_\nu(\epsilon_B)$ for all $\nu$. In other
words, as $\epsilon$ decreases, the function $R_\nu(\epsilon)$
monotonically increases. Finally, we note that $R_\nu(\epsilon)$
converges pointwise to $R_\nu$ as $\epsilon$ goes to zero. At this
point, we invoke Dini's theorem, which states that a sequence of
monotonically increasing, continuous and pointwise convergent
functions on a compact space are uniformly convergent. Therefore,
there is a positive number $\epsilon_4$ such that whenever
$\epsilon<\epsilon_4$, we have $R_\nu(\epsilon)>M/2$.

For such values of $\epsilon$, the inequalities (\ref{cond:A}) and
(\ref{cond:B}) are satisfied for all $\nu\in[1/2,2]$. Moreover,
the inequality (\ref{cond:C}) is also satisfied, as
$R_1(\epsilon)=\sigma(x)-\sigma(z^\epsilon)> M/2>0$.

As a result, if
$\epsilon<\min(\epsilon_1,\epsilon_2,\epsilon_3,\epsilon_4)$, then
the sequences $x$ and $z^\epsilon$ satisfies all the inequalities
(\ref{cond:A}-\ref{cond:C}). The proof of case B enables us to
conclude that $x\prec_T z^\epsilon$. Finally, by $z^\epsilon \prec y$ we
reach to the desired result $x\prec_T y$.

\section{Discussion and Conclusion}
\label{sec:conclusions}

A set of necessary and sufficient conditions are given for the
trumping relation. The conditions involve a continuous variable,
but they are easy to verify for concrete examples.

Conditions (\ref{cond:A}-\ref{cond:C}) can be easily adopted to
the sequences in the closure of $T(y)$, where
\begin{equation}
  T(y)=\{x: x\prec_T y\}\quad,
\end{equation}
is the set of sequences trumped by $y$. In that case, if
$x\in\overline{T(y)}$ then the conditions
(\ref{cond:A}-\ref{cond:C}) must be satisfied but with strict
inequalities replaced with non-strict ones.

If $x\in\overline{T(y)}$, but $x$ is not trumped by $y$, it means
that no catalyst can achieve the conversion of $x$ into $y$ with
probability 1, but it is possible to find a sequence of catalysts
(with growingly large Schmidt numbers) such that the conversion
probability is made to approach 1. Interestingly, this property is
also shared by states that are far from the boundary of $T(y)$.
Consider the example,
\begin{eqnarray}
  x &=& \left(\frac{2}{9},\frac{3}{9},\frac{4}{9}\right)\quad,\\
  y &=& \left(\frac{1}{5},\frac{2}{5},\frac{2}{5}\right)\quad.
\end{eqnarray}
As $x_3^\uparrow>y_3^\uparrow$, $x$ is not in the closure of
$T(y)$. However, it can be verified that $A_\nu(x)>A_\nu(y)$ for
all $\nu<1$. This then implies that, any given probability less
than 1 can be achieved by a suitable catalyst in the conversion of
$x$ into $y$.\cite{SuperTrump} However, the elements of
$\overline{T(y)}$ satisfy an additional property, i.e., they can
be catalytically converted \emph{with unit probability} to another
state only slightly different from $y$. It is puzzling to see that
this property is not shared by the pair $x$, $y$ given in the
example above.

Once it is understood that catalysis is possible, the problem of
finding a suitable catalyst can in principle be solved by going
backwards along the proofs. Although possible solutions of the
problem posed in the Lemma in Section \ref{sec:sec1} can be found
by the well-established procedures of linear programming, carrying
out the whole procedure for realistic cases might be forbidding,
as the degree of the polynomial $\gamma(s)$ and of the sought for
polynomial $a(s)$ might be very large. However, the method used in
the proof of the Lemma can used to place an upper bound on the
degree of $a(s)$ (but not on the Schmidt number). This also
suggests a conjecture that the complex roots, $\nu$, of the
equation $A_\nu(x)=A_\nu(y)$, and their closeness to the real line
could be used for estimating the minimum amount of resources the
catalysts should have.

\appendix
\section{Proof of Theorem 2}
\label{app:A}

The most troublesome part of the proof of Theorem 2 is the
neighborhood of $\nu=1$. This part can be handled with the
following theorem.

\textbf{Theorem 3.} For any positive sequence $x$ and any given
$\delta>0$, there is a positive number $\epsilon$ such that, for
any $\bar{x}$ with $\sum_i \bar{x}_i=\sum x_i$, and
$D(x;\bar{x})\le\epsilon$ we have
\begin{equation}
  e^{-\delta \vert\nu-1\vert}\le\frac{A_\nu(\bar{x})}{A_\nu(x)}
  \le e^{\delta
  \vert\nu-1\vert}\quad\forall\nu\in\left[1/2,2\right]\quad.
\end{equation}

\textit{Proof of Theorem 3.} Without loss of generality, it is
supposed that $x=x^\uparrow$. Let $S_\nu(x)=\sum_i x_i^\nu$ be the
$\nu$th power sum and $K_\nu$ be defined as
\begin{equation}
  K_\nu =\ln\frac{S_\nu(\bar{x})}{S_\nu(x)}\quad.
\end{equation}
Note that $K_1=0$. We place the following bound on the absolute
value of $\nu$ derivative of $K_\nu$,
\begin{eqnarray}
  \left\vert\frac{dK_\nu}{d\nu}\right\vert &=&
     \frac{1}{S_\nu(x)S_\nu(\bar{x})} \left\vert \sum_{ij}\bar{x}_i^\nu x_j^\nu\ln\frac{\bar{x}_i}{x_j} \right\vert \\
  &\le&\epsilon+\frac{1}{S_\nu(x)S_\nu(\bar{x})} \left\vert \sum_{ij}\bar{x}_i^\nu x_j^\nu\ln\frac{x_i}{x_j} \right\vert \\
  &\le& \epsilon+\frac{1}{S_\nu(x)S_\nu(\bar{x})}
    \left\vert \sum_{i>j}(\bar{x}_i^\nu x_j^\nu-\bar{x}_j^\nu x_i^\nu)\ln\frac{x_i}{x_j} \right\vert\nonumber
\end{eqnarray}
Since, for $i>j$ we have $x_i\ge x_j$, all of the logarithmic
terms are non-negative in the expression above. As a result, for
any positive $\nu$,
\begin{eqnarray}
  \left\vert\frac{dK_\nu}{d\nu}\right\vert
  &\le& \epsilon+\frac{e^{\nu\epsilon}-e^{-\nu\epsilon}}{S_\nu(x)S_\nu(\bar{x})}
     \sum_{i>j}x_i^\nu x_j^\nu \ln\frac{x_i}{x_j}  \\
  &\le& \epsilon+\frac{e^{\nu\epsilon}-e^{-\nu\epsilon}}{2S_\nu(x)S_\nu(\bar{x})}
     \sum_{i,j}x_i^\nu x_j^\nu \left\vert\ln\frac{x_i}{x_j} \right\vert \\
  &\le& \epsilon+\frac{e^{\nu\epsilon}-e^{-\nu\epsilon}}{2}\frac{S_\nu(x)}{S_\nu(\bar{x})}
    \ln\frac{x_n}{x_1}
\end{eqnarray}
Finally, we can apply $S_\nu(\bar{x})\ge e^{-\nu\epsilon}S_\nu(x)$
to the last line which gives
\begin{eqnarray}
  \left\vert\frac{dK_\nu}{d\nu}\right\vert
  &\le& \epsilon+\frac{e^{2\nu\epsilon}-1}{2}\ln\frac{x_n}{x_1} \\
  &\le& \epsilon+\frac{e^{4\epsilon}-1}{2}\ln\frac{x_n}{x_1}
\end{eqnarray}
where the last inequality is valid for all $0<\nu\le2$. Note that
the right-hand side of the last expression has zero limit as
$\epsilon\rightarrow0$. This enables us to choose the value of
$\epsilon$ so small that the right hand side is less than
$\delta/2$. In other words, $\vert dK_\nu/d\nu\vert\le\delta/2$.

Next, we express $K_\nu$ as
\begin{equation}
  K_\nu=\int_1^\nu \frac{dK_\nu}{d\nu} d\nu\quad.
\end{equation}
The inequality above then implies that
\begin{equation}
  \vert K_\nu \vert \le \frac{1}{2}\delta\vert\nu-1\vert  \quad.
\end{equation}
Finally, considering only the values of $\nu$ in the interval
$[1/2,2]$, we have
\begin{equation}
 \left\vert \ln \frac{A_\nu(\bar{x})}{A_\nu(x)}\right\vert=\frac{\vert K_\nu \vert}{\nu} \le \frac{2\delta\vert\nu-1\vert}{\nu}\le \delta\vert\nu-1\vert  \quad,
\end{equation}
which is the desired result.$\Box$

\textit{Proof of Theorem 2:} Without loss of generality suppose
that $x$ and $y$ are normalized, i.e., $\sum x_i=\sum y_i=1$.
Since the minimum and maximum values of the these sequences are
different by the assumptions of the theorem, the strict
inequalities are valid at the infinities, i.e., $A_{-\infty}(x)>
A_{-\infty}(y)$ and $A_{\infty}(x)<A_{\infty}(y)$. Let $G_\nu$ be
defined as
\begin{equation}
G_\nu = \ln\frac{A_\nu(y)}{A_\nu(x)}\quad.
\end{equation}
By the inequalities (\ref{cond:A}-\ref{cond:C}), we have $G_\nu<0$
for all $\nu<1$ and $G_\nu>0$ for all $\nu>1$. At infinities
$G_\nu$ approaches to non-zero limits. Moreover, the derivative of
$G_\nu$ at $\nu=1$ is
\begin{equation}
  G^\prime_1 = \sigma(x)-\sigma(y)>0\quad.
\end{equation}
Therefore, both of the following quantities are strictly positive,
\begin{eqnarray}
 B &=&  \min_{\nu\in[-\infty,1/2]\cup [2,\infty]} \vert G_\nu\vert  \quad,\\
 M &=& \min_{\nu\in[1/2,2]} \frac{G_\nu}{\nu-1}  \quad.
\end{eqnarray}

By Theorem 3, there are numbers $\epsilon_1$ and $\epsilon_2$ such
that $D(x;\bar{x})\le\epsilon_1$ implies that $\vert \ln
A_\nu(\bar{x})/A_\nu(x)\vert\le M\vert\nu-1\vert/3$ and
$D(y;\bar{y})\le\epsilon_2$ implies that $\vert \ln
A_\nu(\bar{y})/A_\nu(y)\vert\le M\vert\nu-1\vert/3$. We choose
$\epsilon=\min(\epsilon_1,\epsilon_2,B/3)$.

Let $\bar{x}$ and $\bar{y}$ be arbitrary sequences such that $\sum
\bar{x}_i=\sum \bar{y}_i=1$, $D(x;\bar{x})\le\epsilon$ and
$D(y;\bar{y})\le\epsilon$. Let
\begin{equation}
  \bar{G}_\nu =\ln\frac{A_\nu(\bar{y})}{A_\nu(\bar{x})}
  =G_\nu  +  \ln\frac{A_\nu(\bar{y})}{A_\nu(x)}  +  \ln\frac{A_\nu(x)}{A_\nu(\bar{x})}\quad.
\end{equation}
Our purpose is to show that $\bar{G}_\nu$ satisfies the desired
properties, i.e., it is negative for $\nu<1$, positive for $\nu>1$
and has a simple zero at $\nu=1$. Note that
$D(x;\bar{x})\le\epsilon$ implies that $\vert \ln
A_\nu(x)/A_\nu(\bar{x})\vert\le\epsilon$ for all $\nu$.

We consider the following ranges of $\nu$ values separately,
\begin{itemize}
\item[(a)] For $\nu\le1/2$, we have $\bar{G}_\nu\le
-B+2\epsilon\le -B/3<0$.

\item[(b)] For $\nu\ge2$, we have $\bar{G}_\nu \ge B-2\epsilon\ge
B/3>0$.

\item[(c)] For $\nu\in[1/2,2]$ we have
\begin{equation}
\frac{\bar{G}_\nu}{\nu-1} \ge M-\frac{2M}{3}>0\quad.
\end{equation}
As a result, $\bar{G}_\nu$ satisfies the desired properties in
this interval as well.
\end{itemize}

Finally, for the inequality (\ref{cond:C}), we note that for
$\nu\in[1/2,1)$, we have
\begin{equation}
  \frac{\bar{G}_\nu}{\nu-1} \ge \frac{M}{3}\quad.
\end{equation}
Taking $\nu\rightarrow1$ limit gives the derivative of
$\bar{G}_\nu$ which is
\begin{equation}
  \bar{G}_1^\prime = \sigma(\bar{x})-\sigma(\bar{y}) \ge \frac{M}{3}\quad.
\end{equation}
This completes the proof of Theorem 2.

\end{document}